# Bayesian Parameter Inference and Uncertainty-Informed Sensitivity Analysis in a 0D Cardiovascular Model for Intraoperative Hypotension

Jan-Niklas Thiel[a,*], Marko Zlicar[b], Ulrich Steinseifer[a], Borut Kirn[c,†], Michael Neidlin[a,†]

**Affiliation:**
a) Cardiovascular Engineering, Applied Medical Engineering, Medical Faculty, RWTH Aachen University, Aachen, Germany
b) Clinical Department of Anaesthesiology and Intensive Therapy, University Medical Centre Ljubljana, Ljubljana, Slovenia
c) Institute of Physiology, Medical Faculty, University of Ljubljana, Ljubljana, Slovenia

† These authors contributed equally to this work.

**\* Correspondence:**
Name: Jan-Niklas Thiel
Address: Institute of Applied Medical Engineering, Forckenbeckstr. 55
52074 Aachen, Germany
Email address: thiel@ame.rwth-aachen.de

**Funding:**
Funded by the Deutsche Forschungsgemeinschaft (DFG, German Research Foundation) SPP2014 "Towards the Artificial Lung" (project number: 447746988) and "Young Investigator Grant" as part of SPP2014.

**Conflict of interest:**
All the authors have nothing to disclose.

**Ethics statement:**
-

**Authors' contributions:**
**Jan-Niklas Thiel:** Conceptualization, Data curation, Formal analysis, Funding acquisition, Investigation, Methodology, Software, Validation, Visualization, Writing –original draft. **Marko Zlicar:** Conceptualization, Methodology, Writing – review & editing. **Ulrich Steinseifer:** Funding acquisition, Project administration, Resources, Supervision, Writing – review & editing. **Borut Kirn:** Conceptualization, Methodology, Project administration, Resources, Writing – review & editing. **Michael Neidlin:** Conceptualization, Funding acquisition, Methodology, Project administration, Supervision, Writing – review & editing.

**Abstract**

Computational cardiovascular models are promising tools for clinical decision support, particularly in complex conditions, such as intraoperative hypotension (IOH). IOH arises from different mechanisms, making treatment selection non-trivial. Patient-specific predictions require calibration, typically performed using deterministic approaches prone to parameter non-identifiability and lacking uncertainty quantification, hindering clinical translation. Consequently, Bayesian approaches are needed that facilitate parameter inference, sensitivity analysis, and uncertainty quantification in cardiovascular models.

We utilize Bayesian Markov chain Monte Carlo (MCMC) to estimate parameter distributions of a cardiovascular lumped parameter model (LPM) across different IOH scenarios. We demonstrate parameter non-uniqueness and its impact on sensitivity indices. We improve parameter reliability by incorporating clinical knowledge and measurement uncertainties. We enable continual learning of the model using sequential parameter updating as new patient data become available. We introduce an uncertainty-aware sensitivity analysis and compare it with a deterministic approach.

Deterministic calibration yielded many local solutions for IOH, with notably different sensitivities. MCMC distinguished different hypotension scenarios, such as those induced by impaired contractility or hypovolemia. Parameter uncertainty decreased by about 70% with additional data, and by up to 94% with sequential updating. Propagating uncertainties from MCMC through sensitivity analysis provided tighter credible intervals, resulting in more stable parameter rankings than the deterministic approach. The Bayesian approach revealed differences in model behavior and treatment suggestions across patient conditions, highlighting the potential to inform therapy planning.

Combining Bayesian inference with sequential updating and sensitivity analysis improves the reliability and identifiability of parameter estimates, enhancing the clinical utility of LPMs for therapy guidance.

# 1 Introduction

Computer models offer the potential to support clinical decision-making in complex, highly patient-specific therapies. One example is the treatment of intraoperative hypotension (IOH). IOH is characterized by a significant decrease in mean arterial pressure (MAP) and may be caused by various factors, including severe vasodilation (vasoplegia), hypovolemia due blood loss, or reduced ventricular contractility [1, 2]. Much research has focused on developing algorithms to predict IOH. Work from Hatib et al. [3] has developed one such algorithm, the Hypotension Prediction Index (HPI), that is already applied in clinical practice [4–6]. The HPI predicts the risk of IOH occurring within the next 10 minutes using a score ranging from 0 to 100. However, it does not directly support clinical decision-making regarding the best intervention strategy once IOH occurs in these patients. This is crucial because it is often difficult to know a priori which option will effectively treat the underlying cause and increase the MAP to a safe level [7–12]. Treatment options include vasopressors to induce vasoconstriction, inotropes, used to increase cardiac contractility, and intravenous fluid boluses to increase overall systemic pressure. Therefore, it is desirable to provide clinical decision support that estimates the effect of each treatment option for individual patients with IOH. Computational models can deliver this information [13].

Any computational tool intended for clinical use must be efficient and robust and require minimal user input to reduce misuse and inter-operator variability. Lumped parameter models (LPMs) meet these requirements and have been applied to various biological systems, for example, the human cardiovascular system [14–17]. These models divide the human body into an arbitrary number of 0D compartments represented by Windkessel elements of different orders. This results in a set of ordinary differential equations and unknown model parameters, which can vary in magnitude depending on the patient and pathology. The advantage of LPMs is that they are derived from physiological mechanisms, making their parameters clinically interpretable. However, calibrating them is often challenging. Typically, this is done using a deterministic approach, which generates point-wise parameter values [18–21]. Calibrating complex, multi-parameter LPMs with deterministic approaches is challenging because solutions are often non-unique due to highly nonlinear relationships between model parameters and outputs, resulting in parameter non-identifiability [22–24]. This is a crucial point for robust clinical decision-support systems because multiple, physiologically distinct parameter combinations can reproduce the same clinical measurement, leading to different model behavior and unreliable predictions. This fundamental problem can be addressed in various ways. One approach is to rank parameters and reduce the number of free parameters included in the calibration using a sensitivity analysis, while retaining the overall model structure [25]. As an alternative approach, uncertainties can be explicitly considered in the calibration, and the resulting predictions and credible intervals must lie within clinically meaningful thresholds for the model's intended context of use. The model can be considered sufficiently calibrated if the uncertainty of the task-specific prediction - which may refer to specific treatment indications or risks, or be used for diagnosis and disease classification - lies within the defined limits [26].

Unlike classical deterministic approaches, which yield point estimates, Bayesian inference determines statistical distributions, called posterior distributions, of parameter values that have the highest likelihood of producing certain observed data. It enables the quantification of various sources of uncertainty in the measurement data and their propagation to model outputs by sampling from the estimated posterior distributions and performing forward simulations. A sampler iteratively updates user-defined prior distributions to obtain those posteriors. These priors can contain different amounts of information and can be defined based on previous experiments or clinical (expert) knowledge. When little to no information is available, they can be weakly or non-informative [27]. For instance, priors can be defined as a uniform distribution within known physiological ranges or as a normal distribution considering a specific diagnosis. For instance, if a patient presents with reduced cardiac function, a higher likelihood of decreased ventricular contractility can be assumed. Posterior distributions can also be updated sequentially as new measurement data become available by using posteriors from a previous time instance as new priors. This avoids having to recalibrate a Bayesian model from scratch, saving computational resources and time [28, 29]. One such algorithm is the Markov chain Monte Carlo (MCMC) method, which has been used in several studies, including those related to cardiovascular LPMs [26, 30–33]. A practical limitation of Bayesian approaches is that they require a large number of model evaluations for parameter inference, which can be computationally infeasible for large, complex models, particularly in clinical practice. Consequently, reduced order models (ROMs) are often used to accelerate sampling [33, 34].

The problem of parameter identifiability can be addressed by conducting a sensitivity analysis with subsequent reduction of free parameters. This involves varying the model parameters within defined ranges and quantifying their effect on the model outputs. The most prominent approach is the Sobol method, a variance-based global sensitivity analysis (GSA) [35–37]. The resulting sensitivity indices can be used to rank the parameters and define subsets for the calibration process that exclude correlated parameters [25, 38–40]. Relating the most sensitive parameters of a computational model, as identified by a GSA, to their physiological effects might have the potential to support clinical decision-making. Essentially, a GSA could improve understanding of how the human body dynamically responds to specific therapies and medications. In most studies, though, GSA is performed at a single, fixed model state, and a nominal parameter set is perturbed within small, defined ranges [20, 26, 41]. However, in mechanistic models that exhibit non-identifiability, meaning that multiple parameter sets explain the measurements equally well, the sensitivity indices and resulting rankings or therapy suggestions are non-unique and vary across the possible parameter space. In consequence, to ensure a robust calibration procedure across patients and pathologies, a GSA must provide stable rankings valid across the relevant output range (e.g., patients with normotension, hypotension, and hypertension). To achieve this, the GSA can be performed for multiple nominal parameters or the perturbation applied to a single chosen model state can be increased, in order to avoid redesigning the calibration procedure for each patient. To obtain reliable therapy suggestions when using a GSA to support clinical decisions, uncertainty quantification (UQ) must be integrated to obtain distributions and credible intervals for Sobol indices [39, 42–44]. This can be achieved through upstream Bayesian inference. To the author's knowledge, no study with a clinical focus has yet considered uncertainties in sensitivity indices, and only a few have performed a sophisticated GSA across multiple model states. This idea is similar to the approach used by Eriksson et al. to analyze dynamical intracellular pathway models [45].

One clinical pathophysiological event that would benefit from supporting therapy decisions with sensitivity indices is IOH. There is non-uniqueness in interventions (vasopressors, inotropes, and fluid boluses) that can potentially lead to the same desired clinical outcome (an increase in MAP). Because parameters of cardiovascular LPMs (resistances, compliances, elastances) have clear physiological meanings, they can be mapped to the available treatments for IOH. We hypothesize that, for a given patient, the parameters with the highest sensitivity can identify the most promising treatment option for hemodynamic resuscitation.

First, this study aims to highlight the effect of parameter non-uniqueness, which comes from deterministic parameter inference approaches, on sensitivities derived from classical GSA. Second, the study aims to develop a methodology to integrate UQ into GSA to produce uncertainty-aware sensitivity indices with credible intervals. Third, it aims to evaluate the reliability of suggestions derived from these two approaches in the context of clinical decision-support. This contribution promotes the translation of computational models as tools for prediction and therapy guidance. For this purpose, we use a cardiovascular LPM to model the clinical scenario of IOH. In the classical approach, we identify all parameter combinations that yield the same IOH output (mimicking point estimates from deterministic calibration) and perform a GSA for each. In the uncertainty-aware approach, we train a ROM and perform Bayesian inference using MCMC to obtain posterior distributions that reproduce the event of IOH. These posteriors are then propagated through GSA to obtain distributions and credible intervals of the sensitivity indices.

## 2 Materials and methods
### 2.1 Model structure and deterministic parameter inference

An LPM of the cardiovascular system was taken from previous work. The model consists of 12 compartments describing the hemodynamics of the heart and the systemic and pulmonary circulation, as shown in Figure 1. Cardiac function is modeled using an elastance-driven pressure-volume relationship, and vessels are represented by resistance, compliance, and inertance elements. More detailed information about its implementation and all parameter values used can be found in previous work [46]. Model inputs are the systolic elastance of the left ventricle $Emax_{lv}$, the resistance of the systemic microcirculation $Rmc$, the change in the patient's total blood volume TBV, and the heart rate bpm. The model computes pressures and flows and reported outputs are cardiac output CO, mean arterial pressure MAP, and pulse pressure PP.

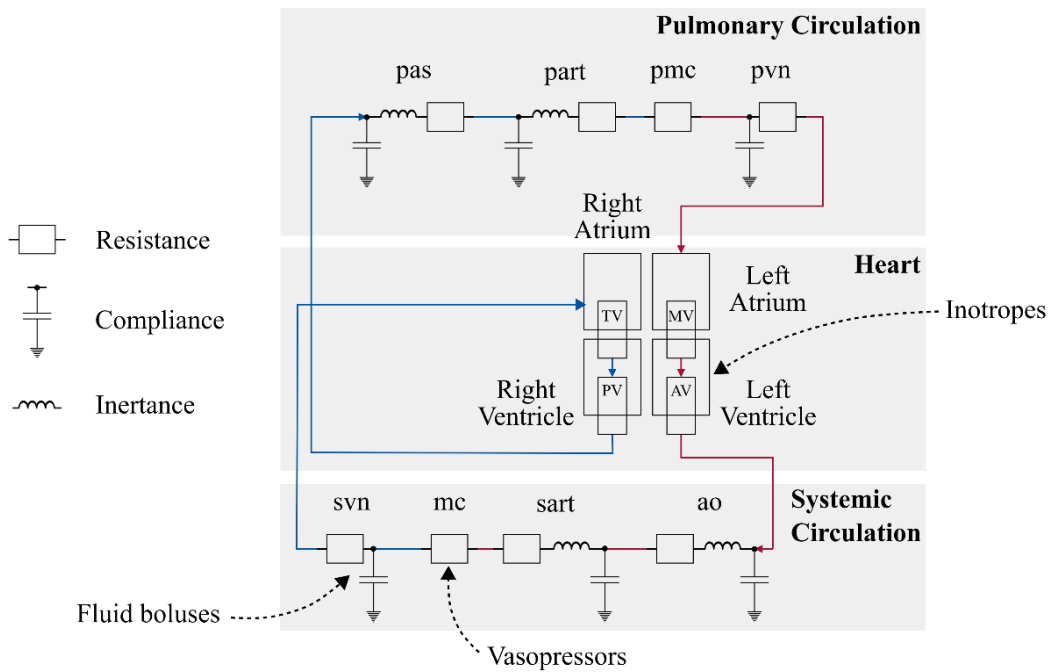

Figure 1: Overview of the cardiovascular lumped parameter model, taken from previous work [46]. Linked changes representing different treatments for intraoperative hypotension are shown with dashed arrows.

The simulation time was set to 100 s, with an adaptive time step of 5×10⁻⁴ s and the ODE system was solved using the Dormand Prince algorithm. The last two cycles were evaluated. There are three therapy options for IOH. The first is inotropes, which increase ventricular contractility and may elevate both CO and MAP. In contrast, vasopressors increase systemic vascular resistance through vessel constriction, which tends to decrease CO while increasing MAP. Administering an additional fluid bolus elevates MAP and can affect CO differently depending on whether vasoplegia is present (CO increases) or cardiac function is impaired (CO decreases). These treatment options were linked to specific model parameters. The effect of inotropes was mimicked by increasing the systolic elastance of the left ventricle Emaxlv. Vessel constriction due to vasopressors was represented by increasing the resistance of the systemic microcirculation Rmc, the largest compartment of the arterial system. Because additional fluid increases intravascular pressure, this effect was modeled by adding a pressure source term to the differential equation of the systemic venous compartment (svn, see Figure 1). Instantaneous infusion or loss of fluid induces a pressure change dP$_{inf}$, implemented by relating the change in the patient's total blood volume TBV to the compliance of the systemic venous compartment Csvn and a pulse duration t$_{inf}$. This source term is applied once the simulation reaches a stationary state:

$$dP_{svn} = dP_{svn} + dP_{inf}, \tag{1}$$

$$dP_{inf} = \frac{TBV}{Csvn * t_{inf}}. \tag{2}$$

The pulse duration t$_{inf}$ is set to 0.01 s. Ramping up and down is implemented using a smoothed Heaviside function to avoid discontinuities, which improves adaptive time-stepping and the stability of the gradient calculation.

To illustrate the issue of non-uniqueness in deterministic approaches for parameter inference, the LPM was evaluated at varying Emaxlv, Rmc, and TBV values within the ranges presented in Table 1. Grid sampling was used to generate 343 samples. The results for non-sampling points were determined using cubic spline interpolation. Results for MAP and CO were then visualized using heatmaps, with one parameter fixed to its median value at a time. Additionally, contour lines were created for target values of MAP = 65 mmHg and CO = 3 L min⁻¹, as well as for 10% deviations from these values.

Table 1: List of parameters and their ranges used in this study.

| Parameter | Range | Type |
|---|---|---|
| Systolic elastance of the left ventricle Emaxlv in mmHg mL⁻¹ | [0.1, 5.0] | Input |
| Resistance of the systemic microcirculation Rmc in mmHg s mL⁻¹ | [0.1, 2.0] | Input |
| Change in total blood volume TBV in mL | [-500, 500] | Input |
| Heart rate bpm in min⁻¹ | [50, 120] | Input |
| Cardiac output in L min⁻¹ | - | Output |
| Mean arterial pressure in mmHg | - | Output |
| Pulse pressure in mmHg | - | Output |

## 2.2 Global sensitivity analysis

A variance-based GSA was utilized to quantify the effects of changes in Emaxlv, Rmc and TBV on MAP and CO, which are proxies for the treatment of IOH. The resulting sensitivity indices range from zero to one and higher values indicate greater influence of the corresponding model parameter on the clinical marker. More details on the implementation can be found in previous work that uses the *SALib* package [46, 47]. For each sample point, a GSA was conducted with parameter perturbations that emulate the changes induced by each of the clinical treatment options. Perturbation ranges were derived from the literature and are listed in Table 2. A sample size of $N = 512$ was used together with $D = 3$ cardiovascular system parameters, resulting in $N(2D + 2) = 4,096$ model evaluations and converged total order indices. Saltelli's extension of the Sobol' sequence was used for sampling.

Table 2: List of parameter perturbations used for the global sensitivity analysis.

| Parameter | Perturbation |
|---|---|
| Systolic elastance of the left ventricle Emaxlv | ±20 % [48, 49] |
| Resistance of the systemic microcirculation Rmc | ±0.2 mmHg s mL⁻¹ [50, 51] |
| Change in total blood volume TBV | ±140 mL [52] |

The results from the GSA provided parameter sensitivities for the ranges listed in Table 1. These were illustrated using heatmaps, in which the colors are weighted by each parameter's individual contribution to the sum of $S_T$. This provides insight into which parameter is most sensitive to MAP and CO at any given point in the parameter space.

### 2.3 Parameter inference using Bayesian approach

Since Bayesian inference with MCMC requires a large number of model evaluations, we trained ROMs to make model calibration efficient and computationally feasible for clinical application. These substitutes for the full order cardiovascular LPM follow:

$$(CO, MAP, PP) = g(Emaxlv, Rmc, TBV, bpm; \beta). \quad (3)$$

As previously described, model inputs are Emaxlv, Rmc, TBV, and BPM, while model outputs consist of CO, MAP, and PP. $\beta_i \in \mathbb{R}$ are model parameters specific to the trained ROM. The open-source Python tool *SASQUATCH*, which was developed in a previous study, was used to train Non-Intrusive Polynomial Chaos Expansion (NIPCE) models of different orders. In this context, $\beta_i \in \mathbb{R}$ are the estimated expansion coefficients for each polynomial basis term. Cheng et al. and our previous study provide more details on the formulation and theoretical background of NIPCE models [47, 53]. 5,000 data points were generated by sampling from the ranges presented in Table 1 using Latin hypercube sampling and by evaluating the cardiovascular LPM for these samples. Training and testing were performed using k-fold cross-validation with ten equally sized random splits. The performance of the models was compared using the $R^2$ score and the mean absolute percentage error (MAPE), shown in Figure S-1 in the supplements. For the following analysis, we used the NIPCE model of order three.

MCMC is a collection of algorithms used to produce dependent samples that approximate the joint distribution of model parameters for a computational model describing patient-specific or clinical cohort data, while accounting for uncertainty. They create Markov chains that start from arbitrary parameter values, which, after an initial burn-in, eventually converge to a target posterior distribution. Prior distributions, which reflect initial beliefs and incorporate clinical knowledge, are combined with a likelihood, which links the model to measurement data and accounts for uncertainties, such as noise, outliers, and other measurement errors. Combining both, these define the posterior (i.e., the updated belief), which is the target distribution that MCMC aims to sample from. Common algorithms include Metropolis-Hastings, Gibbs sampling, slice sampling, and Hamiltonian Monte Carlo (HMC), including the No-U-Turn Sampler (NUTS). These algorithms iteratively draw many dependent samples and produce chains that explore the parameter space, attempting to move to points that are more probable than the existing ones, i.e., points in higher-density regions. Samples obtained from the initial warm-up phase are discarded. The retained draws provide distributions for each parameter, rather than single-point estimates (as in deterministic inference), from which one can report means, medians, and credible intervals. Accounting for uncertainties contributes to enhancing the reliability of parameter estimates and subsequent model predictions, a critical aspect for clinical translation. A more thorough description of the theory and concepts behind MCMC can be found in the comprehensive works of Speagle [27], van de Schoot [54], et al., and Hartig, et al. [55].

The cardiovascular model parameters Emaxlv and Rmc were estimated using the MCMC implementation of the open-source Python package *PyMC*. The gradient-based NUTS algorithm was utilized and the change in total blood volume (TBV) and the heart rate (bpm) were treated as clinically measurable inputs and kept fixed. Four chains were run, each using 5,000 samples after an initial burn-in of 1,000 samples, resulting in 20,000 post-warm-up draws in total. A target acceptance rate of 0.95 was used. The convergence of each MCMC run was assessed using various standard diagnostics. First, the rank-normalized split $\hat{R}$ was required to be less than 1.01, and the effective sample size (ESS) to be greater than $100 \times$ number of chains for all parameters. These diagnostics were visualized in Figure S-2. Second, the posterior distributions were examined alongside trace and rank plots, as shown in Figure S-3. The trace plots for each chain must demonstrate random fluctuations around a stable horizontal band, and all chains need to overlap well to indicate sufficient mixing. Furthermore, it is necessary that the samples span the entire range and do not get stuck at specific values. For the rank plots, the distribution of ranks must be uniform across chains. This is true when all rank bins have approximately the same height, meaning all chains are exploring the same distribution equally. Lastly, autocorrelation is quantified and visualized in Figure S-4 and was required to decay rapidly toward zero.

In total, four different clinical scenarios were investigated using MCMC, as illustrated in Figure 2, to address questions relevant to clinical translation:

1. Can MCMC produce parameter posteriors that distinguish different patient conditions?

2. Are there different ways to apply MCMC, and how do they affect the estimated parameter distributions?

3. How can uncertainty in the posteriors be reduced?

More information on the patient's condition, the prior distributions used, and the abbreviations for each scenario can be found in Table 3. Corner plots were used to illustrate the parameter distributions for Emaxlv and Rmc for all four scenarios.

To investigate whether MCMC can identify different patient conditions, patients under normotension and hypotension, both with and without blood loss, were examined. In the first scenario, the patient had normal arterial pressures and cardiac output, and the measurements used for this normotensive scenario (Normo) were derived from synthetic data points. To generate these synthetic data, the following target values were specified: CO = [5.0, 5.5, 6.0] L min$^{-1}$, MAP = [87, 90, 93] mmHg, PP = [39, 40, 41] mmHg, and bpm = [65, 70, 75] min$^{-1}$. Around each target value, normal distributions were created with standard deviations taken from the measurement uncertainties listed in Table 4. From these distributions, 100 samples were drawn, randomly shuffled and concatenated. This resulted in a large sample of synthetic clinical data for each clinical marker, avoiding cross-marker pairing. Six samples were randomly selected from this synthetic data for further analysis, reflecting the limited availability (sparsity) of routine clinical measurements. These samples are listed in Table 5 and were used for the MCMC analysis. To account for measurement uncertainty, the standard deviation of the measurement error for flows and pressures was taken from Table 4. For the remaining scenarios, it was assumed that the patient's blood pressure and cardiac output were dropping and would eventually reach the critical state of hypotension (Hypo), with a MAP of 65 mmHg and a CO of 3 L min$^{-1}$. In the Hypo A scenario, it was assumed that there was no significant blood volume depletion. In contrast, scenario Hypo B mimics a state of hypotension combined with a mild form of hypovolemia resulting from a blood loss of 333 mL. Independent uniform prior distributions were used for the scenario of normotension (Normo) and Hypo A Independent with the bounds listed in Table 1, reflecting the absence of prior clinical knowledge. In contrast, Hypo A and Hypo B Sequential used the posterior from the normotensive scenario as the initial prior, and then performed ten sequential calibrations, each time taking the posterior from the previous step as the next prior while linearly decreasing the observations for CO and MAP from normotension to hypotension. The main idea with this sequential updating is to approximate the marginal distribution of the last posterior and fit a multivariate normal distribution to capture dependencies. This is then used with its mean and covariance to define the joint prior for the next calibration step.

Table 3: Clinical scenarios investigated using parameter inference with MCMC.

| Clinical scenario | Prior knowledge | Prior distribution | Abbreviation |
| --- | --- | --- | --- |
| Normotension | No | Uniform | Normo |
| Hypotension | No | Uniform | Hypo A Independent |
| Hypotension | Yes | Posterior from Normo | Hypo A Sequential |
| Hypotension | Yes | Posterior from Normo | Hypo B Sequential |

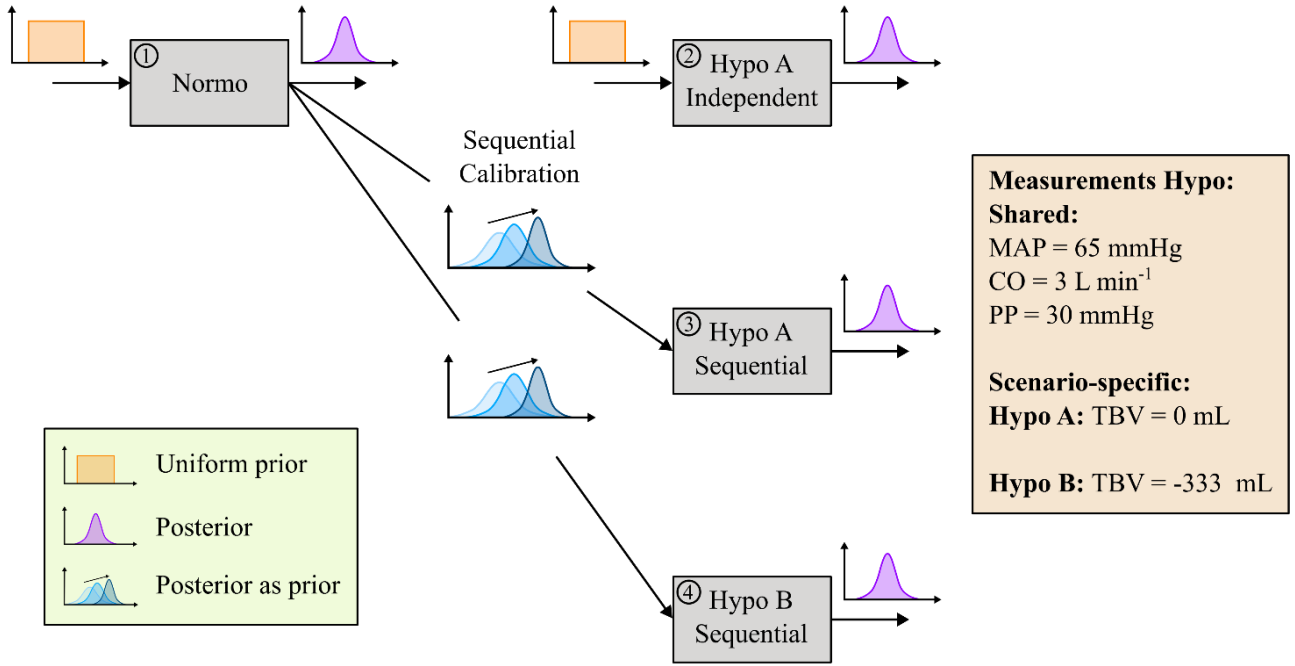

Figure 2: Overview of the different clinical scenarios used for parameter inference with MCMC: (1) Normo and (2) Hypo A Independent are calibrated using uniform priors. (3) Hypo A and (4) Hypo B Sequential use posteriors from (1) Normo as initial priors. Measurements used to define the likelihood of the hypotension scenarios are shown in the orange box.

Table 4: List of measurement uncertainties applied to the observed data used in the MCMC method.

| Observation | Uncertainty |
|---|---|
| Pressures in mmHg | $0.025 \times \text{median(pressures)} + 2$ [56, 57] |
| Flows in L min$^{-1}$ | $0.05 \times \text{median(flows)} + 0.2$ [58, 59] |

Table 5: Synthetic measurement data of a patient with normotension used for the parameter inference with MCMC.

| CO in L min$^{-1}$ | MAP in mmHg | PP in mmHg | TBV in mL | bpm in min$^{-1}$ |
|---|---|---|---|---|
| 5.8, 5.3, 4.8, 6.3, 6.0, 5.8 | 89.1, 80.7, 93.9, 96.0, 91.8, 97.1 | 38.2, 37.1, 37.3, 39.4, 43.9, 47.8 | 0 | 66.7, 65.6, 71.1, 71.0, 75.1, 72.9 |

To compare different approaches of applying MCMC, we compare the standard single-run setup with uniform priors (Hypo A Independent) with a Bayesian sequential updating (Hypo A Sequential), using iterative parameter calibration as described above.

To quantify the effect of the number of patient data points on the uncertainty of the estimated parameter distributions, the normotensive scenario (Normo) was run using 3-30 samples drawn from the synthetic dataset introduced earlier. For each case, MCMC was run, and the coefficients of variation (CV) were calculated following

$$CV = \frac{\sigma}{\mu} \qquad (4)$$

which relates the standard deviation σ to the mean value μ of each model parameter's posterior distribution. Lastly, the influence of sequential parameter calibration on the posterior uncertainty of each model parameter was quantified by varying the number of calibration steps between normotension to hypotension from 0 to 75 in Hypo A Sequential. CV was plotted for the posteriors of each model parameter.

## 2.4 Sensitivity analysis for deterministic and Bayesian approaches

Sensitivity indices derived from estimates of the Bayesian MCMC approach were compared to sensitivities computed from the local solutions of the deterministic approach. For that purpose, 1,000 samples were drawn from the parameter distributions estimated in Hypo A Sequential. A GSA was performed for each sample, with the perturbations listed in Table 2. Using a sample size of 8,000 resulted in $N(2D + 2) = 64,000$ evaluations of the previously trained ROM. As a result, uncertainties in the estimation of model parameters were projected onto the derived total order sensitivity indices $S_T$. The resulting distributions of $S_T$ were compared to the sensitivities obtained in the deterministic approach for TBV = 0 mL using box plots for each parameter and output.

Similarly, distributions of sensitivities were calculated for Hypo B Sequential and compared to the values obtained for Hypo A Sequential. This comparison aims to determine whether these two different patient states influence the magnitude of $S_T$ and its corresponding uncertainty.

## 3 Results
### 3.1 Local solutions of deterministic approach and implications on sensitivity analysis

Figure 3 shows heatmaps of MAP (top row) and CO (bottom row) for different values of Emaxlv, Rmc, and TBV. The solid contour line indicates the targets of MAP = 65 mmHg and CO = 3 L min$^{-1}$, while the dashed contours show ±10% variation from these targets (target areas). The black dots represent candidate single-point estimates within these ranges. Altogether, 28 distinct points were identified within the MAP and CO target areas. The target areas for MAP cover the entire range of parameter values. The same is true for the CO target area when Rmc is kept constant. However, when TBV is fixed at 0 mL (first column), the CO target area is limited to low Emaxlv values below 1 mmHg mL$^{-1}$. With a constant Emaxlv of 1.7 mmHg mL$^{-1}$ (third column), a CO of 3 L min$^{-1}$ occurs only with blood losses greater than 250 mL. In general, all results for MAP and CO cover the entire range of physiological values of Emaxlv, Rmc and TBV as reported in the literature. Increasing Emaxlv, Rmc or TBV increases MAP, and vice versa. Similarly, a positive correlation exists between Emaxlv and TBV and CO. In contrast, increasing Rmc decreases CO.

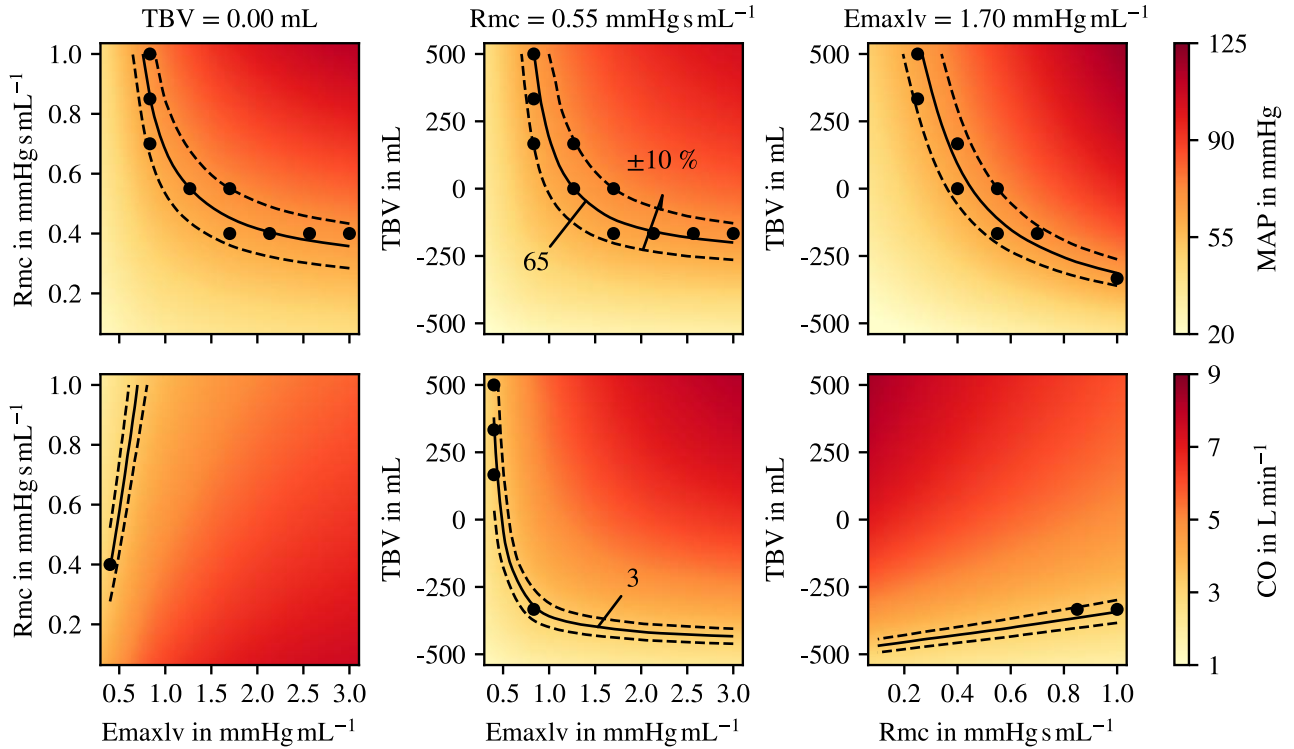

Figure 3: Mean arterial pressure and cardiac output for different values of the systolic elastance of the left ventricle Emaxlv, the resistance of the systemic microcirculation Rmc and the change in total blood volume TBV. Candidate point-wise estimates of the parameters that lie within the target area of hypotension are displayed as black dots.

Figure 4 shows the weighted sum of the total effect $S_T$ of each model parameter, as well as MAP (top row) and CO (bottom row), for different combinations of parameter values. Blue indicates a high sensitivity of CO or MAP to Emaxlv, magenta to Rmc, and yellow to TBV. Black dots again mark the samples within the target area of hypotension. In conditions of hypovolemia with a negative TBV, MAP is primarily influenced by changes in TBV. When TBV is positive, however, the influence shifts toward Rmc, especially at moderate to low Rmc values. The areas where Emaxlv is the most influential parameter are small, i.e., when Rmc is high and Emaxlv is low. A more balanced influence on MAP can be observed when both Rmc and Emaxlv are high. In this region, Rmc and TBV are equally sensitive. Conversely, when both Rmc and TBV are high, Emaxlv and Rmc dominate changes in MAP. CO is most affected by changes in TBV, especially when TBV is negative. Compared to the sensitivities for MAP, there are more regions of balanced sensitivity. Rmc and Emaxlv are only dominant when TBV is positive and either Rmc is very high or Emaxlv is very low. In summary, the contribution of each parameter to the model dynamics is non-unique due to varying parameter sensitivities depending on the chosen parameter combination.

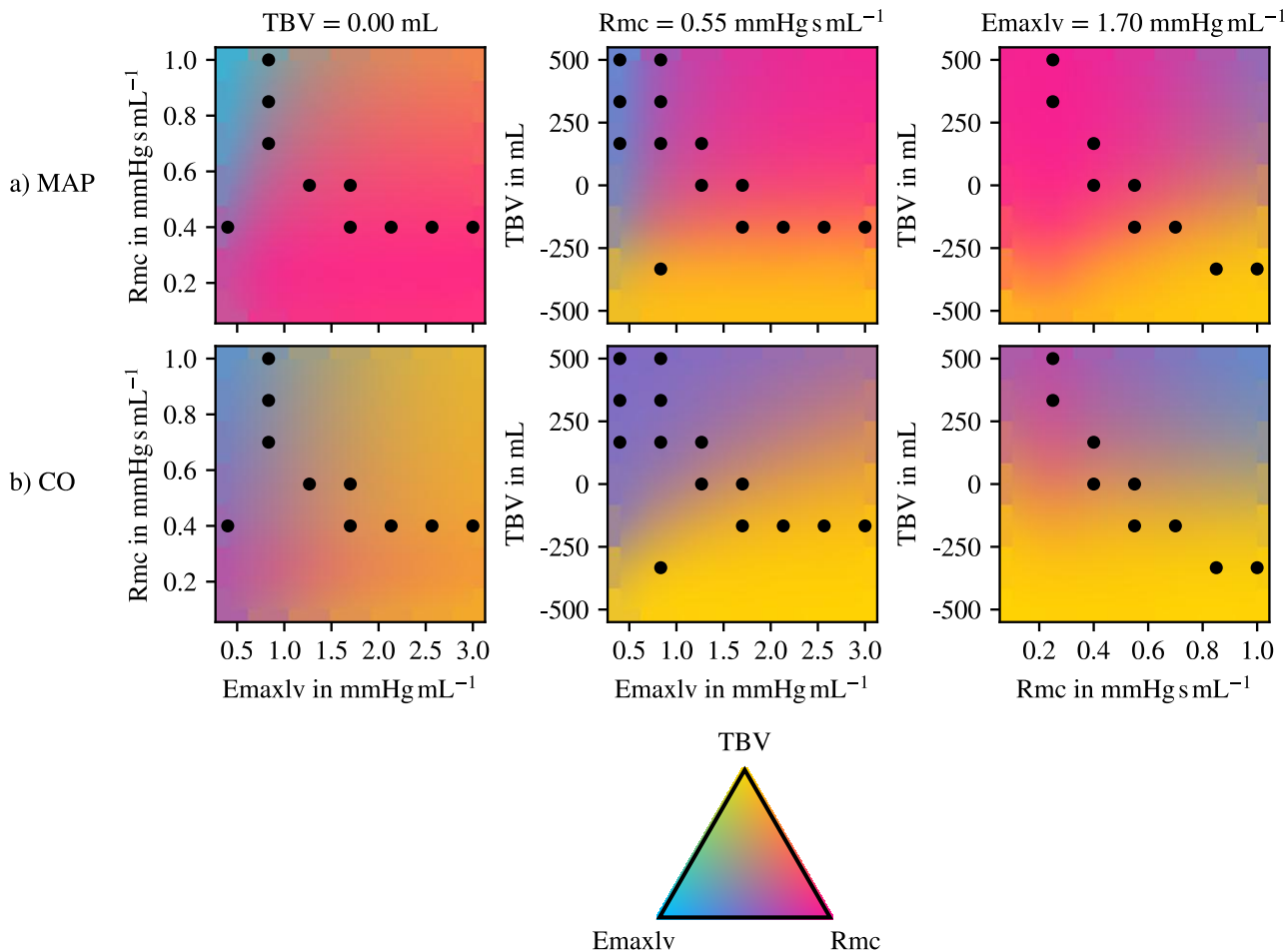

Figure 4: Weighted sum of the total effect $S_T$ of each model parameter for MAP (top row) and CO (bottom row). High sensitivity to Emaxlv illustrated in blue, to Rmc in magenta, and to TBV in yellow.

## 3.2 Parameter estimation using Bayesian approach of MCMC

Figure 5 shows the posterior distributions for Emaxlv and Rmc using corner plots for all clinical scenarios. Their relationship is visualized by combining scatter and contour plots in a joint posterior density plot. Vertical dashed lines mark the 16th, 50th, and 84th percentiles, and the median and standard deviations (68% credible intervals) are reported above the posterior distributions. Given normotensive conditions, MCMC estimates the median of Emaxlv to be 2.02 mmHg mL$^{-1}$ and Rmc to be 0.82 mmHg s mL$^{-1}$, as can be seen in Figure 5-1). No correlation between the parameters can be observed. When the patient suffers from hypotension without any blood loss, displayed in Figure 5-2), the median of Emaxlv decreases to 0.84 mmHg mL$^{-1}$, and Rmc slightly increases to 0.96 mmHg s mL$^{-1}$, indicating an impaired cardiac function. In contrast, when the prior distributions are informed by the MCMC run for normotension (Hypo A Sequential), the median of Emaxlv is slightly higher than in the Hypo A Independent case, with a value of 1.07 mmHg mL$^{-1}$, and Rmc is slightly lower, with a value of 0.92 mmHg s mL$^{-1}$, as shown in Figure 5-3). Additionally, the standard deviations decrease in this scenario. For Hypo B Sequential, in which the patient suffers from hypotension combined with a blood loss of 333 mL, the resulting parameter distributions are visualized in Figure 5-4). The results indicate, that the left ventricular contractility is not impaired, and the median of Emaxlv is estimated to be 2.28 mmHg mL$^{-1}$, with a slight increase of Rmc to 0.92 mmHg s mL$^{-1}$, suggesting the absence of any vasodilation. The evolution of the parameter distributions due to a sequential recalibration, along with their corresponding standard deviations, is illustrated in Figure S-5. It also presents a comparison to both the normotensive condition and Hypo A Independent.

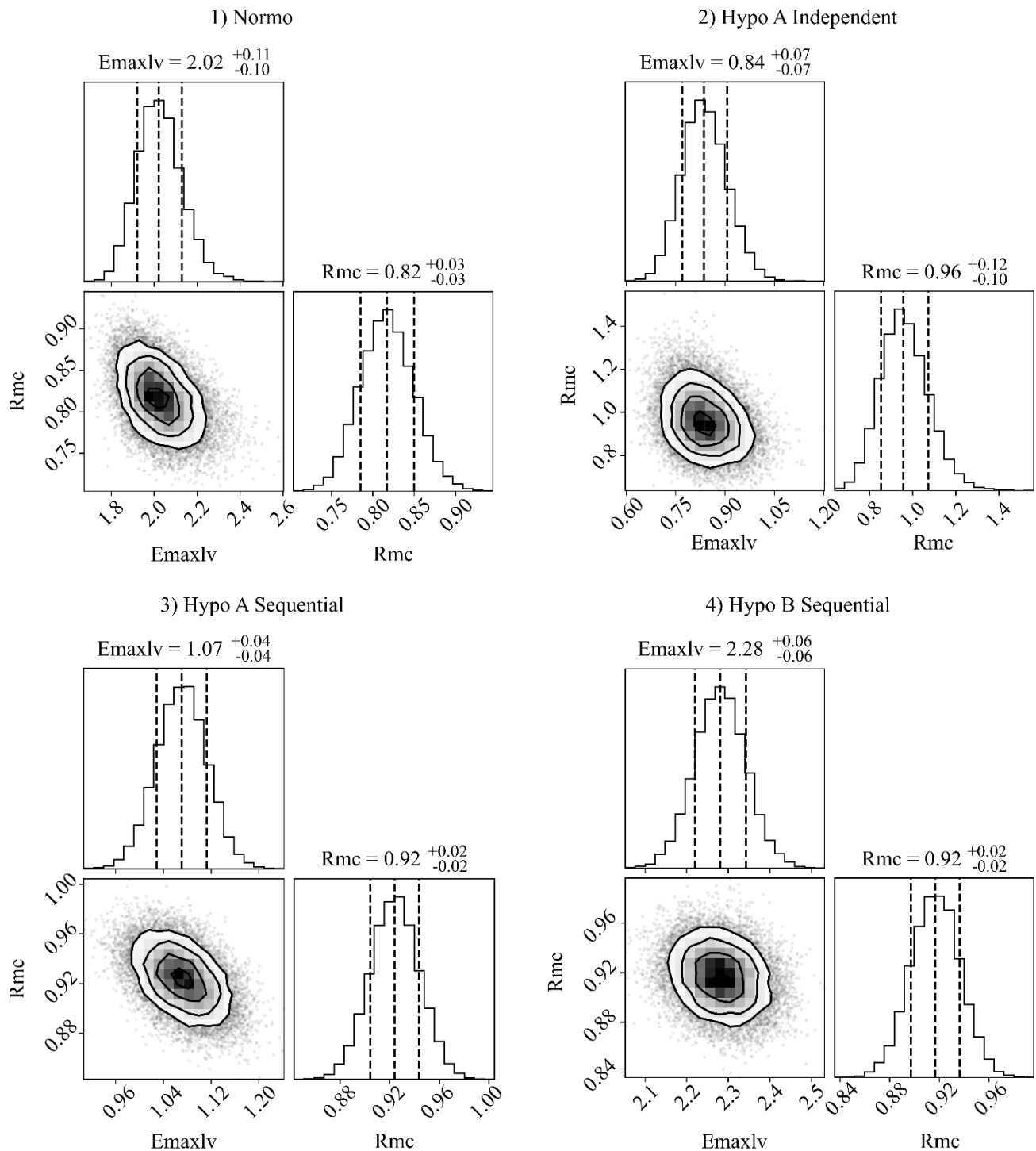

Figure 5: Corner plots showing posterior distributions of each model parameter for the scenarios 1) Normo, 2) Hypo A Independent, 3) Hypo A Sequential and 4) Hypo B Sequential.

Figure 6-a) shows the coefficient of variation of the parameter distributions of Emaxlv and Rmc as a function of the number of clinically measured data points included in the MCMC run for a patient under normotension. The CV decreases as more data points are included and exhibits a monotonic, asymptotic behavior. Using 30 data points reduces the uncertainty of the estimated parameter distributions by 70% and 68%, respectively, for both Emaxlv and Rmc compared to the MCMC run with only 3 data points. Across all tested sample sizes, Emaxlv remains 17-25% more uncertain than Rmc.

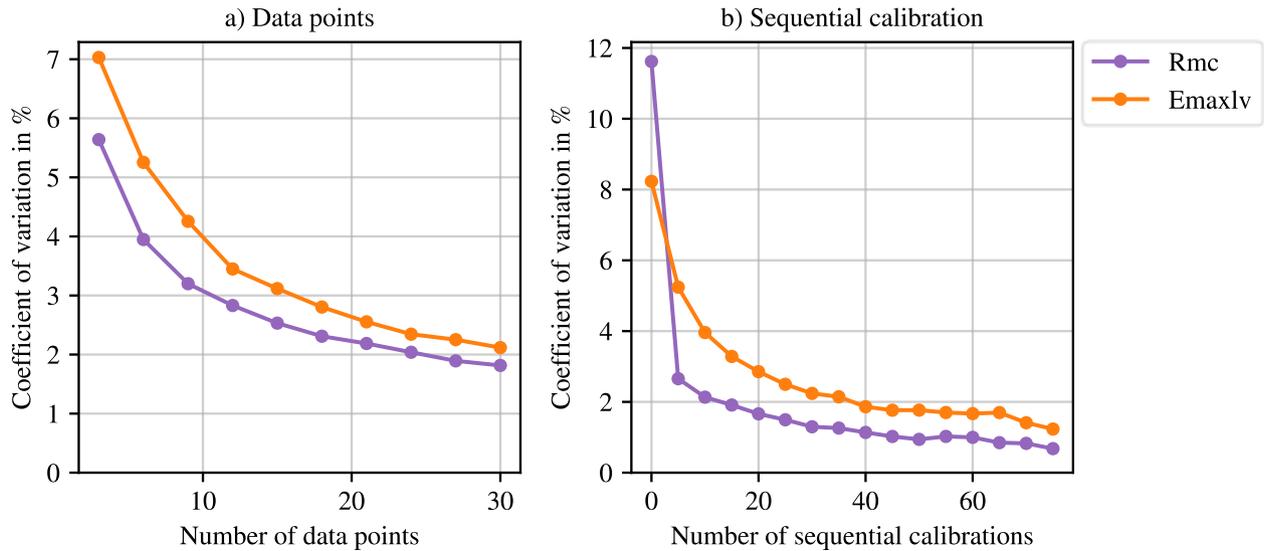

Figure 6: Coefficient of variation for Rmc and Emaxlv, depending on a) the number of data points and b) the number of sequential calibration steps included in the parameter inference using MCMC.

Figure 6-b) illustrates the coefficient of variation of the parameter distributions of Emaxlv and Rmc depending on the number of sequential calibration steps performed during the progression of hypotension. As in the previous analysis, a monotonic, asymptotic behavior can be observed, implying that the more calibration steps performed, the lower the coefficient of variation. Here, the magnitudes of uncertainty for both Emaxlv and Rmc are comparable. Using 75 calibration steps reduces the CV for Emaxlv and Rmc by 85% and 94%, respectively. Again, Emaxlv is more uncertain than Rmc, except for independent calibration without sequential updating.

### 3.3 Uncertainty-aware sensitivity analysis

First, we examine the overall model behavior as expressed by the GSA results, which are obtained using the differing parameter estimates from the deterministic and Bayesian inference approaches. Figure 7 shows box plots of the distributions of the total order sensitivity indices of Emaxlv, Rmc, and TBV for the outputs CO and MAP. It compares the predicted sensitivities of the Bayesian approach, Hypo A Sequential (solid box), with the deterministic approach, for which the target points satisfy TBV = 0 (hatched box). The magnitudes of the sensitivity indices $S_T$ and the ranking of the most influential parameters vary depending on the approach. For the deterministic approach, TBV has the greatest impact on CO, while the effects of Rmc and Emaxlv are moderate and small, respectively. In contrast, the Bayesian approach identifies Rmc as the most influential parameter, with the impact of TBV being moderate and that of Emaxlv being small. Rmc is the most dominant parameter for MAP in both approaches. In general, the deterministic approach tends to predict higher sensitivities for Rmc than the Bayesian approach, but their parameter ranking is the same (Rmc → TBV → Emaxlv). The uncertainties of the sensitivity values are significantly higher in the deterministic approach, resulting in a less stable ranking of the model parameters than in the Bayesian approach.

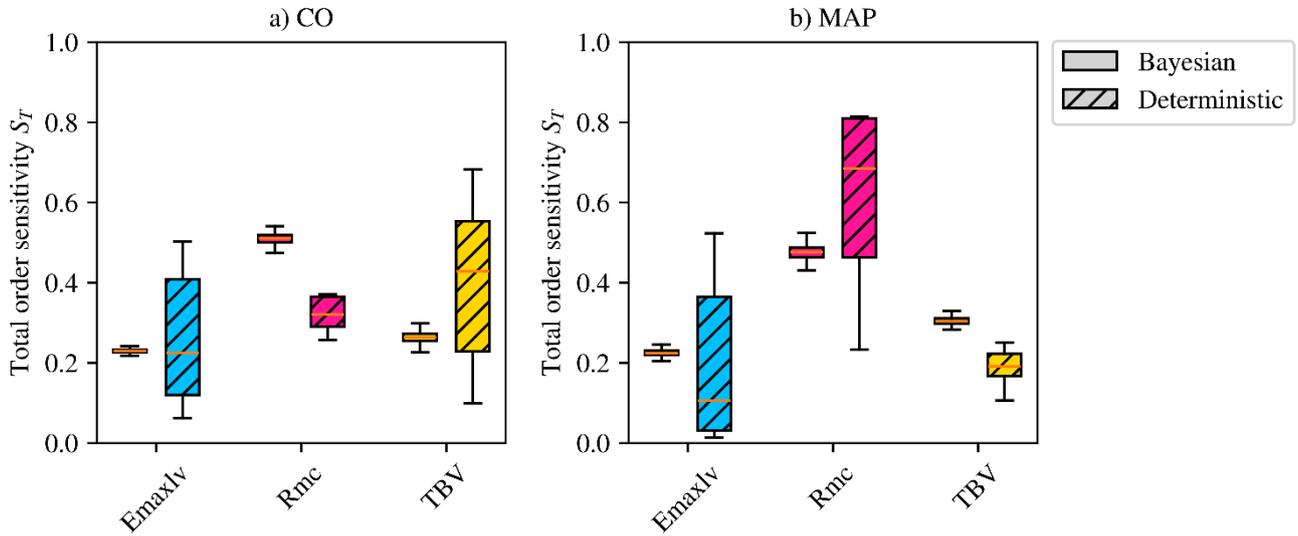

Figure 7: Distributions of the total order sensitivity indices $S_T$ for each model parameter and output for the Bayesian (Hypo A Sequential) and deterministic (target points satisfying TBV = 0 mL) approaches.

Lastly, we assessed the model behavior across patient conditions by running GSAs on posteriors from Hypo A and Hypo B Sequential (hypotension with or without blood loss). Figure 8 shows the distributions of $S_T$, with solid boxes representing Hypo A Sequential and dotted boxes showing Hypo B Sequential. The ranking of the most sensitive model parameters appears to depend on the patient's condition. For Hypo A, both CO and MAP are mainly affected by Rmc, while Emaxlv and TBV have a comparably moderate impact. In contrast, a reverse ordering of the parameter influence can be seen for Hypo B and CO. Emaxlv and TBV both have a significant impact, while Rmc has low $S_T$ values. Changes in MAP in scenario Hypo B are mainly driven by TBV. The sensitivities in Hypo B appear to show a greater variability than in Hypo A, and these uncertainties are generally greater for CO than for MAP. Furthermore, the variability is greater for Emaxlv and TBV than for Rmc.

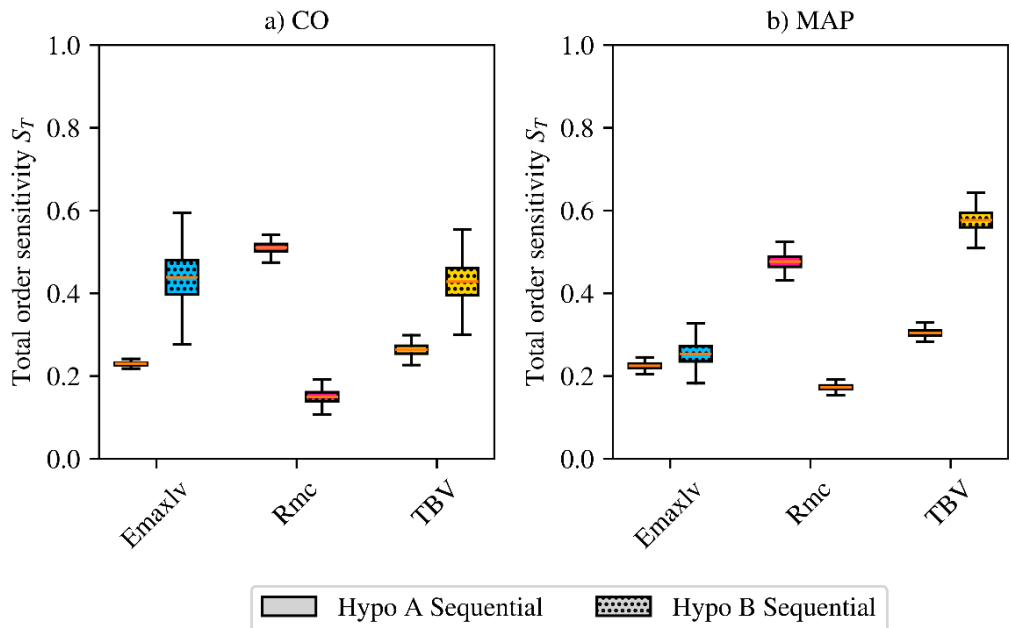

Figure 8: Distributions of the total order sensitivity indices $S_T$ for each model parameter and output for the Bayesian approaches Hypo A and Hypo B that use sequential calibration.

## 4 Discussion

This study aimed to demonstrate the capability of Bayesian parameter inference using MCMC to find parameter distributions for a 0D cardiovascular model in different clinical scenarios of IOH, as well as highlighting the benefits of integrating prior clinical knowledge and uncertainties. Furthermore, the study aimed to demonstrate how this can be used to introduce an uncertainty-aware sensitivity analysis to support clinical decision-making. We successfully developed a strategy to integrate MCMC into computational models for clinical decision-support. This produced parameter distributions that were both clinically interpretable and meaningful. Combining MCMC with GSA allowed us to propagate uncertainty to model sensitivities, mitigating the issue of non-uniqueness and providing valuable new insights relevant for therapy planning. Four main findings were identified:

1. Deterministic calibration results in significantly different model behavior because of non-unique parameter estimates.
2. Bayesian parameter inference can identify unique, physiologically meaningful parameter distributions that describe different patient conditions.
3. Posterior uncertainty decreases as more patient data become available and parameters are updated more frequently using sequential calibration.
4. Using a GSA informed by parameter distributions from MCMC increases confidence in the predicted sensitivities.

In this work, a ROM was trained using a previously developed cardiovascular LPM and was used to apply MCMC to patients with normo- and hypotension. Different amounts of measurement data and calibration steps were tested to quantify their impact on uncertainties in the predicted parameter distributions. These distributions were used in a sensitivity analysis to quantify the effect of every parameter on the clinical markers within each scenario.

From a technical perspective, with a focus on parameter identifiability and inference, model sensitivity, and uncertainty quantification, the results can be summarized as follows: Figure 3 illustrates the issue of non-uniqueness of deterministic approaches for parameter calibration that provide local, point-wise solutions. Even keeping only one of the three parameters fixed resulted in 28 possible solutions for the remaining two free parameters. Consequently, keeping all three parameters free will most likely result in even more possible solutions, further worsening the problem of non-uniqueness. Figure 4 reveals that applying perturbations to different parameter values that lead to the same values of MAP and CO leads to a fundamentally different model behavior. This can have significant implications for interpreting the results of a GSA. As outlined in the introduction, GSA outputs are often used to reduce the number of free parameters included in a calibration to experimental or clinical data. Since overall sensitivities differ greatly depending on the chosen nominal values to which perturbations are applied, this affects decisions about which parameters to select for calibration tasks. In addition, for models that account for both individual patient states and patient trajectories (clinical data acquired over the course of therapy), identifying an "incorrect" parameter set at the beginning of the trajectory invalidates subsequent model predictions. It is worth noting that the model is rather simple, with only three parameters calibrated to two measurements (MAP and CO). Other models, e.g., those from Colunga et al. [30] or Karamolegkos et al. [60], contain more variable parameters, making the issue of non-uniqueness higher dimensional. When using computational models to support clinical decisions, sensitivities that are representative of the wrong patient condition could result in incorrect treatment recommendations.

These problems motivate the use of Bayesian approaches, which can account for uncertainties in estimated sensitivity values. Figure 7 demonstrates the improvement of combining MCMC with GSA over a GSA based on local solutions from deterministic calibrations. Relying on a single-point estimate from a deterministic calibration (one target point in Figure 3) poses the risk of a model representation that does not accurately reflect the patient. Therefore, the derived sensitivities may provide misleading therapy suggestions. A brute-force strategy (deterministic calibrations from many starting points to collect local solutions), as illustrated in Figure 3 and Figure 7, still yields unreliable sensitivities that are unsuitable for clinical decision-making. In contrast, combining MCMC and GSA enables one to propagate posterior means and credible intervals onto model sensitivities, thereby introducing uncertainty quantification of model behavior. This information is incredibly valuable for clinicians to judge the reliability of model predictions. Taking this a step further, Figure 8 shows that this approach works across different patient states and suggests varying treatment modalities accordingly.

This study proposed two methods for integrating Bayesian parameter inference and global sensitivity analysis. First, the independent MCMC approach specifies uniform prior distributions, giving the sampler the greatest flexibility in finding parameter distributions without introducing excessive expert knowledge. While this simplifies the setup, it may impede the convergence of the MCMC chains to the true posterior distributions, as Argus et al. have outlined [26]. This issue is particularly problematic for complex Bayesian models and small sample sizes, as Lambert et al. have demonstrated [61]. The second approach implements sequential calibrations as new patient data becomes available. It uses the parameter distributions from the previous MCMC run as priors for the current iteration. This on-line calibration considers the patient trajectory, which reduces uncertainty in the predicted parameter distributions, as shown in Figure 6-b). However, introducing clinical knowledge in the priors may bias the sampler and hinder it from exploring the entire possible parameter space between each MCMC run. Nevertheless, this approach is more efficient than a full recalibration, particularly for large datasets. It reduces the uncertainty in the estimated parameter distributions, as shown in Figure 6-a), and it has been demonstrated to converge with fewer iterations by DeYoreo et al [28]. In practice, the choice of priors should depend on the clinical context. For a newly admitted patient, as much data as possible should be collected first. Then, an MCMC with uninformative uniform priors should be performed to obtain a baseline representation of the patient. As new measurement data becomes available over time, the posteriors should be updated using the proposed sequential updating approach.

Interpreting the results with respect to the clinical case of intraoperative hypotension, we observe the following: Using a Bayesian approach with MCMC, parameter distributions for different patient conditions were estimated, identifying the different mechanics that cause a drop in MAP and CO. In the baseline scenario of normotension, the estimated Emaxlv is representative of normal cardiac function [15], and Rmc is within the physiological range for the systemic vascular resistance (SVR) [62]. For the Hypo A Sequential scenario (no blood loss and iterative calibration of distributions), the derived parameter distributions suggest an impaired left ventricular contractility, as indicated by a 58% drop in Emaxlv and a 17% increase in SVR from baseline. These observations are consistent with the proposed parameter values of Shi et al. [63] and the described mechanisms of Wijnberge et al. [4]. In contrast, Hypo B Sequential correctly suggests hypovolemia and preserved cardiac function. These results demonstrate that changes in parameter values identified through MCMC align with clinical observations. This highlights the strength of mechanistic models with physiologically interpretable parameters. Furthermore, integrating a GSA based on parameter distributions obtained by MCMC into clinical practice appears feasible. Ideally, parameters are calibrated once using all available data (Normo) and then updated as new patient data becomes available (Hypo A Sequential). This approach enables continual learning of the model by retaining information from earlier, potentially non-pathophysiological states. For IOH, this could support the use of more targeted combinations of vasopressors, inotropes, and fluids, which could help reduce the risk of postoperative acute kidney injury [7] or gastrointestinal anastomotic leak [64].

Given that the contractility Emaxlv is significantly reduced and Rmc is slightly increased for Hypo A Sequential, a balanced treatment approach with inotropes and vasopressors is likely an effective therapy option for this scenario. However, a systematic, variance-based GSA reveals that the treatment with vasopressors is the most efficient way to elevate MAP to normal values. Combining both insights may help decide on a therapy that treats the underlying cause and keeps MAP within the normal physiological range, as recommended by Saugel et al. [65]. For Hypo B Sequential, the distributions of $S_T$ suggest a more balanced approach of combining low doses of inotropes with the administration of fluid boluses.

Fixing the change in total blood volume TBV, which represents blood loss in the case of hypovolemia, has certain implications for clinical application. It reduces model complexity and enhances parameter identifiability, improving the performance and efficiency of the Bayesian parameter inference. However, Saugel et al. emphasize that quantifying intraoperative blood loss is often challenging [65]. For this reason, a precise approximation of the patient's preoperative blood volume is essential, along with the continuous monitoring of the intraoperative blood loss using one of the methods described by Lin et al. which should be tailored to the specific clinical scenario encountered [66].

This study has several limitations. First, the Bayesian parameter inference using MCMC was applied to synthetically generated clinical data. This data was created by assuming random Gaussian noise and applying random, independent shuffling. Consequently, the data does not contain physiological dependencies between samples or clinical markers, and it covers only a small range of possible patient states. However, the goal of this study was to demonstrate a methodological proof of concept and there is no reason to assume that real-world clinical data will change the general observations made. The next logical step is applying it to real patient data. Furthermore, the defined measurement uncertainties are based on data sheets of sensors found in the literature. For real clinical applications, these uncertainties must be specific to the sensors and workflows used. Additionally, a ROM based on a relatively simple 0D cardiovascular computational model was used with MCMC to infer model parameter distributions. It consists only of a combination of resistances, compliances, and inertances. It does not include autoregulation, heart-lung interaction, or other complex physiological mechanisms. This has not yet been validated for the specific clinical case of IOH. The general model behavior derived from GSA was only compared with general clinical observations. Therefore, it remains to be shown whether the model can make predictions for real, patient-specific conditions. Although models simplify reality, we believe they could serve as future clinical decision-support tools as long as their uncertainty is rigorously assessed using MCMC and sensitivity analysis.

## 5 Conclusion

Characterizing the behavior of computational physiology models by combining Bayesian inference based on MCMC with sensitivity analysis and incorporating clinical uncertainties can enhance their clinical utility. Our novel approach enables a) determining parameter distributions that reflect different patient states, b) reducing uncertainty in these distributions through sequential updating, and c) performing uncertainty-aware sensitivity analysis with the potential to inform therapy planning. This approach reveals substantial differences in model behavior and derived treatment suggestions across patient conditions. Additionally, the uncertainty in sensitivity estimates is significantly reduced compared to approaches based on classical, deterministic calibrations.

**Data availability**

The basic version of the cardiovascular model can be found at the GitHub link https://github.com/nikithiel/ECLIPSE (without implementation of fluid bolus). The code to perform surrogate model training and testing can be accessed through https://github.com/nikithiel/SASQUATCH.

# 6 Supplements

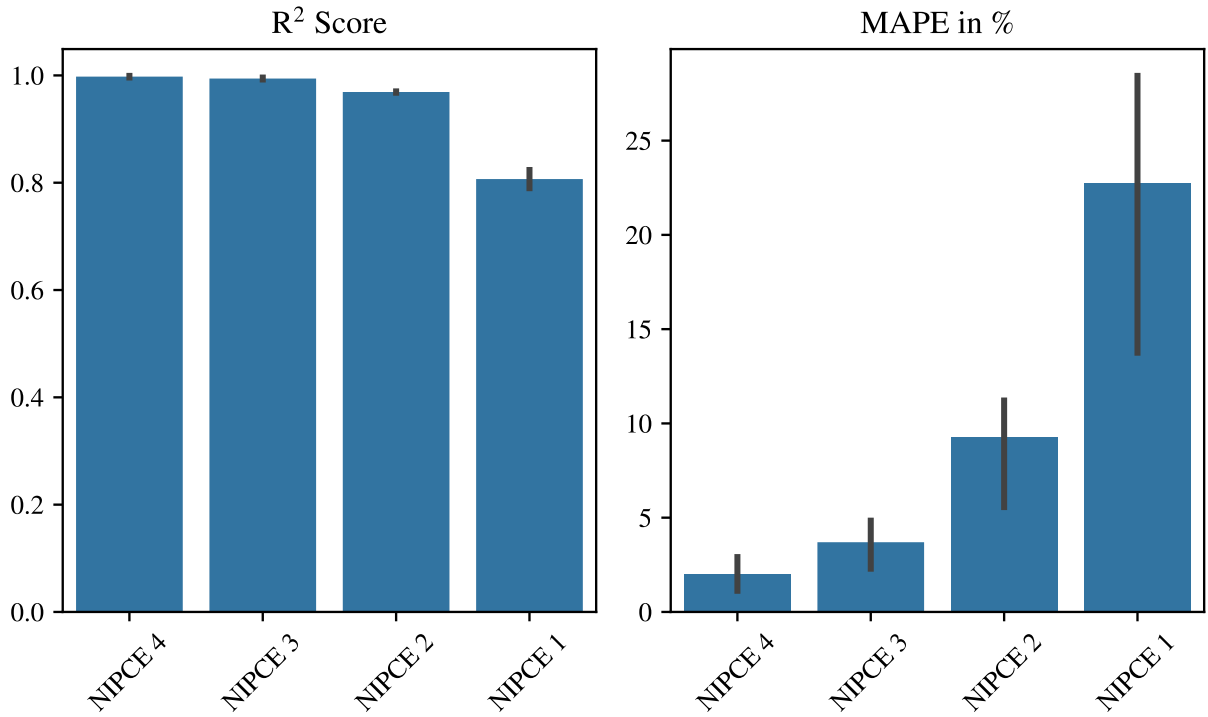

Figure S-1: Performance of NIPCE models of different orders for 5,000 data points created by the cardiovascular LPM.

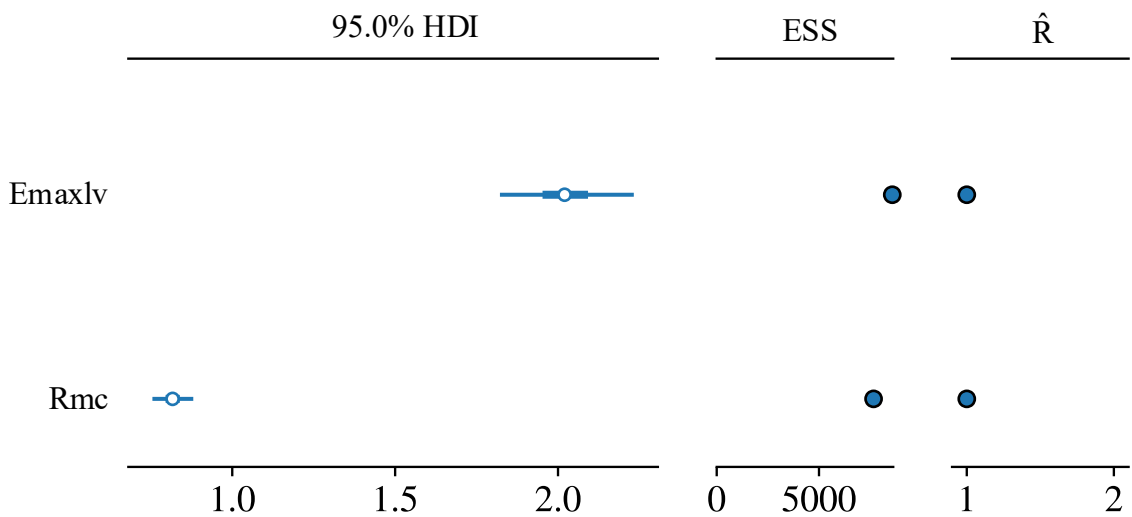

Figure S-2: Forest plot showing the probability of the highest density interval (HDI), the effective sample size (ESS), and the rank-normalized split R-hat ($\hat{R}$) for Emaxlv and Rmc for the normotension scenario.

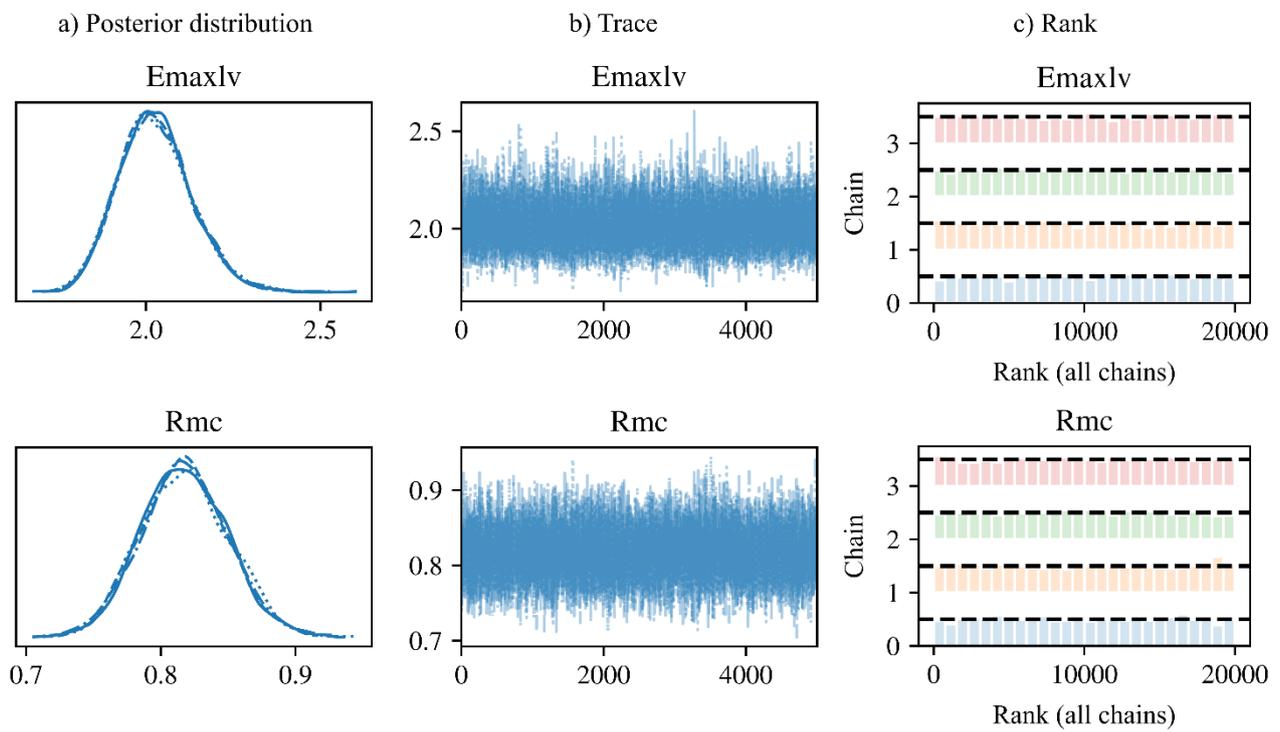

Figure S-3: Posterior distributions in a), trace plots in b) and rank bars in c) for the model parameters Emaxlv and Rmc, obtained for the normotension scenario.

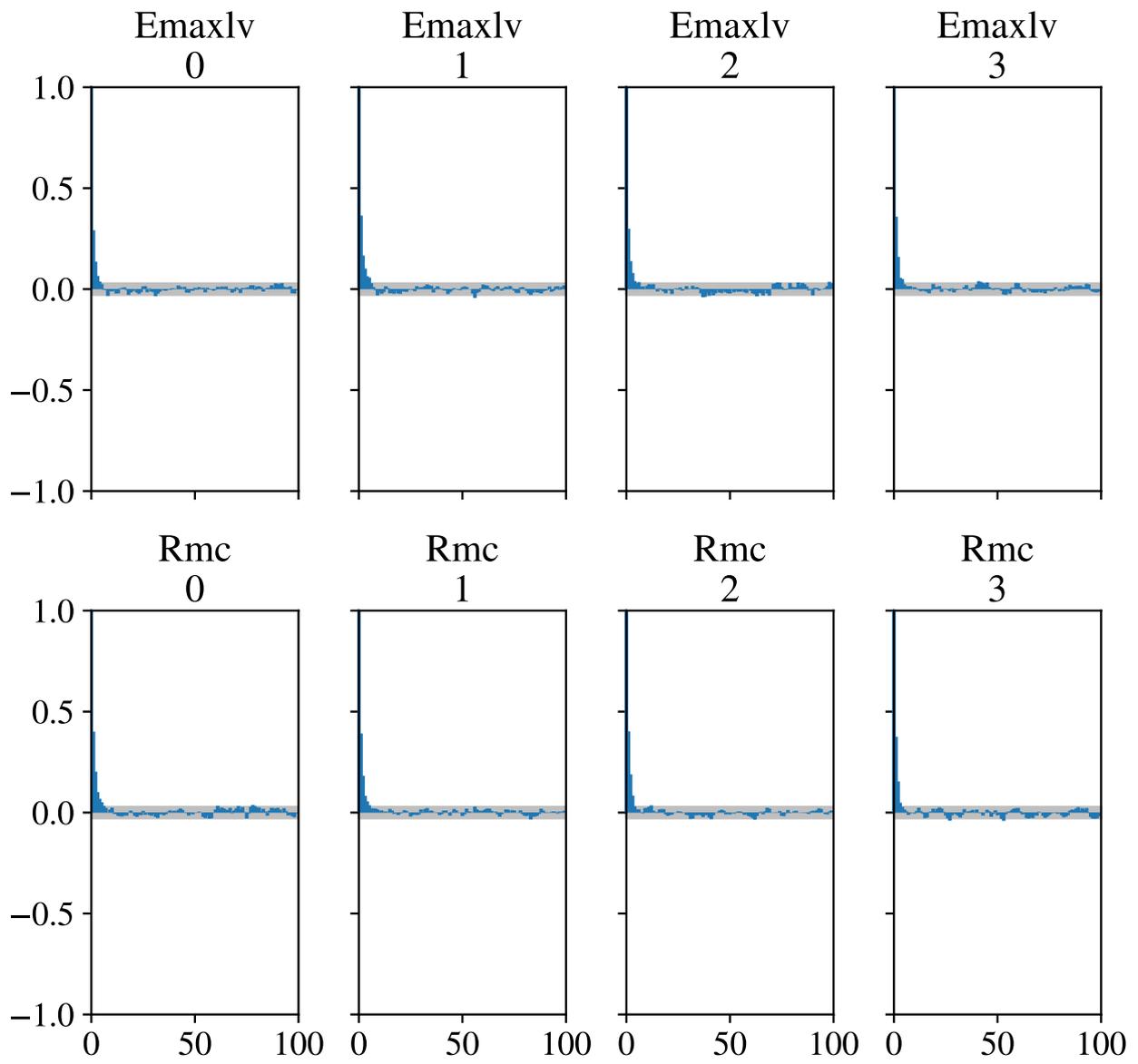

Figure S-4: Bar plots displaying the autocorrelation for each chain and model parameter in the normotension scenario.

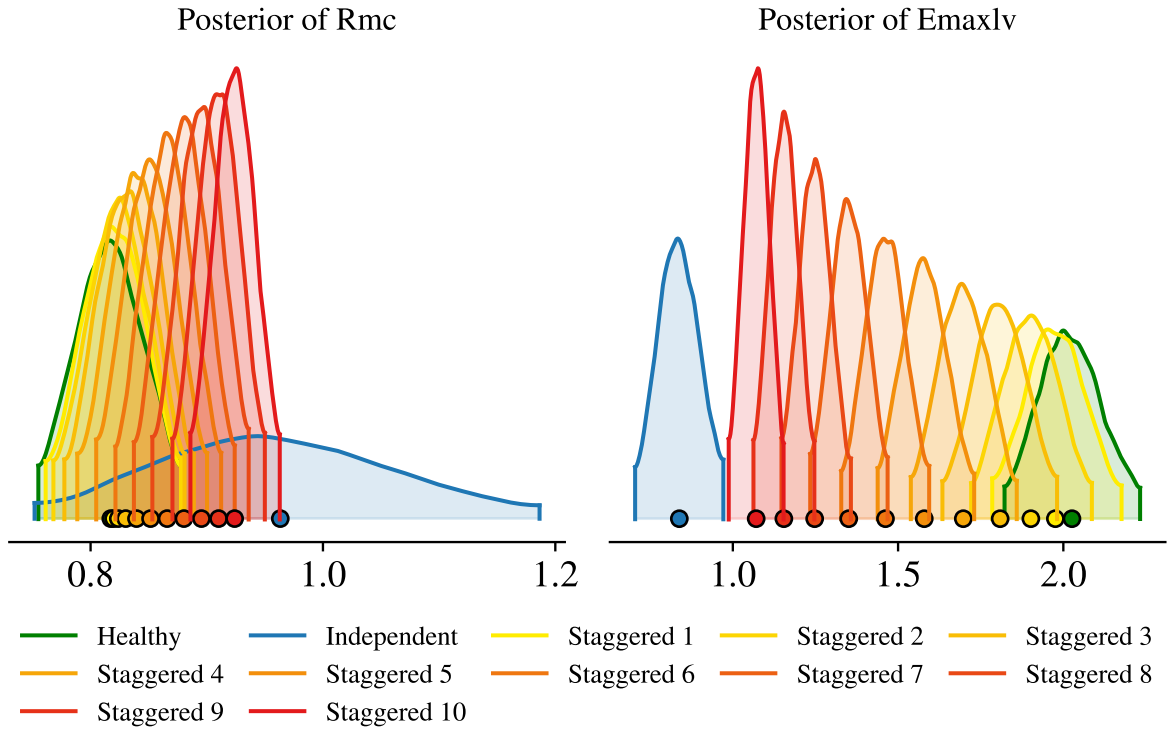

Figure S-5: Trends in the parameter distribution and standard deviation for ten sequential calibration steps from normotension to hypotension (Hypo A Sequential). Results for the patient under normotension are shown in green (Normo). The hypotension scenario with a uniform prior is shown in blue (Hypo A Independent).